\documentstyle[12pt]{article}
\def\BS{{\cal B}}
\def\dl{\delta }
\def\DS{{\cal D}}
\def\FS{{\cal F}}
\def\HS{{\cal H}}

\def\PS{{\cal P}}

\def\SS{{\cal S}}
\def\gm{\gamma }
\def\Gm{\Gamma }

\def\th{\theta }

\def\beq{\begin{equation}}
\def\bt{$\bullet\quad$}
\def\beqa{\begin{eqnarray}}
\def\eeq{\end{equation}}
\def\eeqa{\end{eqnarray}}
\def\hf{{\textstyle {1\over2}}}

\input{epsf}
\begin{document}

\title{Consistent Quantum Realism: A Reply to Bassi and Ghirardi}

\author{Robert B. Griffiths\thanks{Electronic mail: rgrif@cmu.edu}\\
Department of Physics,
Carnegie-Mellon University,\\
Pittsburgh, PA 15213, USA}

\date{Version of 20 Jan. 2000}

\maketitle

\begin{abstract}
	A recent claim by Bassi and Ghirardi that the consistent (decoherent)
histories approach cannot provide a realistic interpretation of quantum theory
is shown to be based upon a misunderstanding of the single-framework rule:
they have replaced the correct rule with a principle which directly
contradicts it. It is their assumptions, not those of the consistent histories
approach, which lead to a logical contradiction.
\end{abstract}

\section{Introduction}
\label{s1}

	The first paper on the consistent histories (CH) interpretation of
quantum theory was published in the Journal of Statistical Physics in 1984
\cite{gr84}.  In the years since then this approach, sometimes called
``decoherent histories'', has been refined and extended in several books and
papers, of which some of the more significant are
\cite{om,gmh,om94,gr96,gr98,om99}.  It provides a realistic
picture of the atomic realm without the need to invoke quantum measurement as a
fundamental principle, and for this reason it can resolve the ``measurement
problem'' \cite{wg63} (there are actually two measurement problems, see
\cite{go99}) which has long beset attempts to place the foundations of quantum
theory on a sound basis, and which probably cannot be dealt with consistently
by traditional methods \cite{pm98}.  Because it resolves various quantum
paradoxes \cite{gr84,gr8799} using an analysis based upon the mathematics
of Hilbert space, the CH approach removes any need to look for alternatives to
standard quantum theory, such as those found in the hidden-variables approach
of de Broglie and Bohm \cite{bohm}, or in the spontaneous
reduction ideas of Ghirardi, Rimini, Weber, and Pearle \cite{grw}.

	There have, to be sure, been a number of criticisms of the CH approach,
and these have proven helpful in constructing improved versions of the
formalism, and better expositions of its physical interpretation.  The most
significant of these criticisms are discussed in \cite{gr98}, and reasons are
given why they do not invalidate CH quantum theory.  This reference provides
the material needed to counter various claims, such as in \cite{gl98},
that the CH approach is logically inconsistent or unsound.  In this connection
it is worth pointing out that even one of its severest critics has admitted
that the CH approach is logically consistent when its rules are properly
followed \cite{kt98r}.

	One of these rules, known as the {\it single framework} (or single
family, single logic, or single set) rule, plays a central role in the CH
approach, as has been repeatedly emphasized in various publications
\cite{om94,gr96,gr98,om99}.  Despite the extensive discussion of this rule in
the CH literature, accompanied by numerous applications to specific problems,
it is still sometimes misunderstood, as in some recent work by Bassi and
Ghirardi \cite{bg99,bg00,bg00b}.  In particular, these authors have claimed, in
an article \cite{bg00} appearing in this Journal, that the CH interpretation of
quantum theory when interpreted in a realistic way using some reasonable
assumptions leads to contradictions in the sense of violating a result of Bell
\cite{bl66}, and Kochen and Specker \cite{ks67}.  On the face of it this seems
rather surprising.  The Bell-Kochen-Specker result shows that a certain type of
hidden-variables approach to quantum theory can lead to a contradiction
because it makes assumptions incompatible with the structure of Hilbert space.
On the other hand, the CH interpretation has been explicitly constructed to
take account of the structure of Hilbert space, and does not rely on hidden
variables in any way.

	Closer examination shows that the Bassi and Ghirardi argument violates
the single-framework rule, and thus the claimed contradiction with
Bell-Kochen-Specker is not a consequence of the principles of the CH approach,
but is instead due to Bassi and Ghirardi's having rejected those principles.
This was pointed out in \cite{gr00a} in response to \cite{bg99}, but since
\cite{bg00} is considerably longer and also somewhat clearer than \cite{bg99},
raises the issue in a somewhat different way, and has appeared in a different
journal, a separate reply to it seems appropriate.  The present article, in
order to be self-contained, contains a certain amount of overlap with
\cite{gr00a}.

	Since the arguments in \cite{bg00} (as in \cite{bg99}) deal entirely
with the Hilbert space description of a system at a single time, most of the
formal machinery of the CH approach---histories, consistency, and assignment of
probabilities by use of the time-dependent Schr\"odinger equation---is not
needed for the following discussion.  The essential point we wish to make is
that a quantum Hilbert space differs in crucial respects from a classical phase
space, and this mathematical difference must be reflected in any valid physical
interpretation of quantum theory.  Importing ``intuitively obvious'' ideas of
classical physics into quantum mechanics without paying adequate attention to
the mathematical structure of the latter, in direct contradiction to the rules
of the CH interpretation, is what has given rise to the contradiction noted by
Bassi and Ghirardi, as we shall show.

	Before dealing with the main issue, we need to indicate the connection
between truth functionals---our name for the homomorphisms denoted by $h$ in
\cite{bg00}---and certain elementary concepts from standard probability theory.
This is done in Sec.~\ref{s2}, and the quantum counterparts of truth
functionals and probability concepts are taken up in Sec.~\ref{s3}, along with
the single-framework rule.  In Sec.~\ref{s4} we show that in their argument
Bassi and Ghirardi have mistakenly replaced the single-framework rule with what
we call the {\it every-framework principle}, which is not only not the same as
the single-framework rule, but stands in direct contradiction to it; for this
reason their argument has basically nothing to do with CH quantum theory.
Section \ref{s5} responds to some other less-important issues in \cite{bg00},
and Sec.~\ref{s6} has a brief conclusion.

\section{Sample Spaces and Truth Functionals}
\label{s2}

	A basic concept in elementary probability theory is that of a {\it
sample space}.  According to Feller \cite{fl68}, the possible outcomes of an
idealized experiment correspond to precisely one and only one point of the
sample space. If a coin is tossed, the sample space consists of two
possibilities, H and T; if a die is rolled, there are six points in the sample
space.  Before the idealized experiment is carried out one does not, in
general, know what the outcome will be, but when it has taken place, {\it one
and only one} of the outcomes actually occurs, i.e., is the {\it true} result
of the experiment. The books on probability theory with which I am familiar do
not seem to employ the terms ``true'' and ``false'', but the way in which they
define a sample space justifies the association of ``true'' with the sample
point that represents the actual outcome, and ``false'' with all the others.
In addition, Feller distinguishes {\it simple} events, the elements of the
sample space, from {\it compound} events which are associated with some subset
of the elements of the sample space.  A compound event is ``true'' if it
contains the point of the sample space which actually occurs, and is ``false''
otherwise.

	Rather than the terminology of ordinary probability theory, \cite{bg00}
uses what I call a {\it truth functional}: a homomorphism (there denoted by
$h$) from a Boolean algebra of events to the set $\{0,1\}$, also thought of as
a Boolean algebra.  In light of the preceding remarks, the connection of a
sample space, and its corresponding event algebra, to a system of truth
functionals can be explained in the following way.  Let $\SS$ be the sample
space, and $\PS$ be some subset of $\SS$, thus a compound event in Feller's
terminology.  The {\it indicator} $P$ of $\PS$ is a function on $\SS$ taking
the value 1 at all points which lie in $\PS$ and 0 at all other points of
$\SS$.  The usual Boolean algebra of subsets of $\SS$ is then isomorphic to a
Boolean algebra $\BS$ of indicator functions in which the greatest element is
the function $I$, equal to 1 at all points of $\SS$; the least element is $0$,
equal to 0 everywhere; the complement of an indicator $P$ is $I-P$; the join
$P\cap Q$ of two indicators is their product $PQ$; and the meet $P\cup Q$ is
the indicator $P+Q-PQ$.

	A truth functional $\th$ is then a function which assigns to each
indicator in the algebra $\BS$ either the value 1 (true) or 0 (false) in a way
which satisfies the following three conditions:
\beq
  \th(I)=1,\quad \th(I-P) = 1-\th(P),\quad \th(PQ) = \th(P) \th(Q).
\label{e.1}
\eeq
It is not hard to show that any such function is necessarily of the form
\beq
  \th_q(P) = P(q) =\cases { 	1 & if $ q \in \PS$,\cr
			0 & if $ q \notin \PS$,}
\label{e.2}
\eeq
where $q$ is some point in the sample space $\SS$.  One should think of $\th_q$
as the the truth functional appropriate for the case in which the sample point
$q$ actually occurs, or is true, since it then assigns the value 1 (true) to
every compound event which contains $q$, and 0 (false) to the ones which do not
contain $q$.  If the sample space is discrete, one can think of $\th_q(P)$ in
probabilistic terms as the conditional probability of $P$ given $q$, assuming
the probability of $q$ is greater than zero, so that the conditional
probability is defined.  It is in this sense, among others, that one can say
that ``true'' is associated with a (conditional) probability of 1, and
``false'' with probability 0, in a probabilistic theory.

	This approach can be employed in classical statistical mechanics in the
following way.  Let $\gm$ be a representative point of the phase space $\Gm$.
A {\it physical property} $P$ of the system corresponds to the subset $\PS$ of
$\Gm$ consisting of those points $\gm$ for which this property is true.  The
corresponding indicator $P(\gm)$ is 1 whenever $\gm$ is in $\PS$, and 0
otherwise.  For example, if $P$ is the property that the total energy of a
one-dimensional harmonic oscillator is less than some constant $E_0$, $\PS$ is
the region inside an appropriate ellipse in the $x,p$ plane ($x$ the position,
$p$ the momentum), and $P(\gm)$ is 1 for $\gm$ inside and 0 for $\gm$ outside
this ellipse.

	Now consider a coarse graining of the phase space into a collection
$\DS$ of $N$ non-overlapping regions or ``cells''.  We can write the identity
indicator $I$ (equal to 1 for all $\gm$) in the form
\beq
  I=\sum^N_{j=1} D_j,
\label{e.3}
\eeq
where $D_j$ is the indicator corresponding to the $j$'th cell. Since the  cells
do not overlap it follows that
\beq
  D_j D_k = \dl_{jk} D_j, 
\label{e.4}
\eeq
consistent with the obvious fact that $I^2 = I$.  The set of $2^N$ indicators
which can be written as
\beq
  P=\sum^N_{j=1} \pi_j D_j,
\label{e.5}
\eeq
with $\pi_j$ is either 0 or 1, form a Boolean algebra $\BS$ using the
definitions of complement, meet, and join introduced earlier.  A truth
functional $\th$ is a function on $\BS$ taking the values 0 or 1 in a way which
satisfies (\ref{e.1}), so it has the form
\beq
  \th_k(P) = \cases { 	1 & if $PD_k = D_k$,\cr
			0 & if $PD_k = 0$,}
\label{e.6}
\eeq
where $D_k$ is one of the elements of (\ref{e.3}).  Note that the collection
$\DS$ of cells constitutes a sample space, because in any given ``experiment''
the phase point $\gm$ representing the system will be in one and only one of
the cells.  The truth functional $\th_k$ corresponds to the case in which the
phase point $\gm$ is somewhere in the cell $\DS_k$; it assigns the value 1 to
all collections of cells whose union contains the phase point, and 0 to all
others.  One can again interpret $\th_k(P)$ as a conditional probability,
assuming that the probability assigned to $\DS_k$ is positive.

	Notice that it is because we are assuming that $P$ is of the form
(\ref{e.5}) that the product $PD_k$ must have one of the two forms on the right
side of (\ref{e.6}): no property of the form (\ref{e.5}) can include part but
not all of some cell $D_k$.  Consequently, (\ref{e.6}) defines a truth
functional for indicators belonging to this particular algebra $\BS$, but not
for all possible properties; in this sense a truth functional is relative to a
particular coarse graining $\DS$, or its Boolean algebra $\BS$. However, in
classical mechanics it is possible to construct a {\it universal truth
functional} which is not limited to a single Boolean algebra, but which will
assign 0 or 1 to {\it any} indicator on the classical phase space in a manner
which satisfies (\ref{e.1}).  To do this, choose some point $\gm_q$ in $\Gm$,
and let
\beq
  \th_q(P)=P(\gm_q).
\label{e.7}
\eeq
That is, $\th_q$ assigns the value 1 to any property which contains the point
$\gm_q$, and 0 to any property which does not contain this point, in agreement
with how one would normally understand ``true'' in a case in which the state of
the system is correctly described by the phase point $\gm_q$.  

\section{Quantum Truth Functionals and the Single Family Rule}
\label{s3}

	The quantum counterpart of a classical phase space is a Hilbert space
$\HS$.  For our purposes it suffices to consider cases in which $\HS$ is of
finite dimension, thus avoiding the mathematical complications of
infinite-dimensional spaces.  Following von Neumann \cite{vn55}, we associate a
quantum property, the counterpart of a set of points in the classical phase
space, with a linear subspace $\PS$ of $\HS$, or the corresponding orthogonal
projection operator or {\it projector} $P$ onto this subspace.  If $I$ is the
identity operator, the negation of a property $P$ corresponds to the projector
$I-P$, and the conjunction $P\land Q$ of two properties corresponds to the
projector $PQ$ {\it in the case in which $P$ and $Q$ commute with each other}.
If $PQ\neq QP$, then neither $PQ$ nor $QP$ is a projector, so there is no
obvious way to define a property corresponding to the conjunction, an issue to
which we shall return.

	The quantum counterpart of a coarse graining of a classical phase space
is a {\it decomposition $\DS$ of the identity}, a collection of mutually
orthogonal projectors $\{D_j\}$ satisfying (\ref{e.4}) whose sum is the
identity, as in (\ref{e.3}).  This decomposition gives rise to a set of
projectors of the form (\ref{e.5}), all of which commute with each other, and
which form a Boolean algebra $\BS$ analogous to the algebra of classical
indicator functions. One can define a quantum truth functional $\th$ on the
elements of $\BS$ in the manner indicated previously: it assigns to every
projector $P$ in $\BS$ a value 0 or 1 in a way which satisfies the three
conditions in (\ref{e.1}).  Once again, any truth functional of this type can
be written in the form (\ref{e.6}) for some $k$, and thus there is a one-to-one
correspondence between truth functionals and the elements of $\DS$, which one
can think of as the quantum version of a sample space.

	The CH approach to quantum theory is ``realistic'' in the sense that it
treats the members of a particular decomposition of the identity, a quantum
sample space, as mutually exclusive possibilities, one and only one of which
occurs, or is true, for a particular physical system at a particular instant of
time, in precisely the same sense as in classical statistical mechanics.  The
difference between quantum and classical physics emerges not when one considers
a single quantum sample space, but when one asks about the relationship between
two or more {\it different} sample spaces.  Here quantum theory is very
different from classical physics because the product of two quantum projectors
$P$ and $Q$ on the same Hilbert space can depend upon the order, and when $PQ$
is unequal to $QP$, neither of these products is a projector.  By contrast, the
product of two indicators on the same classical phase space is always an
indicator, since multiplication is commutative. For example, for a spin-half
particle with components of angular momentum $S_x$, $S_y$, and $S_z$ (in units
of $\hbar$), the projector for the property $S_x=+1/2$, which is $\hf I+S_x$,
does not commute with $\hf I+S_z$, the projector for $S_z=+1/2$.  Consequently,
a key question in quantum theory, with no counterpart in classical physics, is:
How can one make sense out of the conjunction of two quantum properties, such
as $S_x=+1/2$ AND $S_z=+1/2$, when the corresponding projectors do not commute
with each other?

	The answer of the consistent historian is that one {\it cannot} make
sense of $S_x=+1/2$ AND $S_z=+1/2$; it is a {\it meaningless} statement in the
sense that (CH) quantum theory assigns it no meaning.  There are no hidden
variables, and thus there is a one-to-one correspondence between quantum
properties and subspaces of the Hilbert space in CH quantum theory.  Since
every one-dimensional subspace of the two-dimensional Hilbert space $\HS$ of a
spin-half particle corresponds to a spin in a particular direction, there is
no subspace left over which could plausibly represent the property $S_x=+1/2$ AND
$S_z=+1/2$.  To be sure, one might assign to it the zero element of $\HS$, a
zero-dimensional subspace corresponding to the property which is always false
(analogous to the classical indicator which is 0 everywhere). This, in fact,
was the proposal, for this particular situation, of Birkhoff and von Neumann in
their discussion of quantum logic \cite{bvn36}.  It is important to notice the
difference between their approach and the one used in CH.  A proposition which
is meaningful but false is very different from a meaningless proposition: the
negation of a false proposition is a true proposition, whereas the negation of
a meaningless proposition is equally meaningless.  The Birkhoff and von Neumann
approach requires, as they themselves pointed out, a modification of the
ordinary rules of propositional logic, whereas the CH approach does not.%
\footnote{For further remarks on some of these issues, see Sec.~IV~A of
\cite{gr98}.}
However, in CH quantum theory it then becomes necessary to exclude meaningless
talk from meaningful discussions, a task which is not altogether
trivial. 

	Generalizing from this example, the CH approach  requires that a
meaningful probabilistic description of a single single quantum system at a
particular time must employ a {\it single framework}: a single Boolean algebra
of commuting projectors generated, in the sense of (\ref{e.5}), from a specific
decomposition of the identity or quantum sample space.  To be sure, {\it many
alternative descriptions} can be constructed using different decompositions of
the identity; the single-framework rule is certainly not intended to restrain
the imagination of theoretical physicists!  However, {\it combining} results
from {\it different sample spaces} into a single description is
forbidden by the single-framework rule, apart from the following exception.

	Two frameworks involving properties of a single system at a single time
(we are ignoring genuine histories, for which the rules are more complex) are
said to be {\it compatible} provided the two Boolean algebras are parts of a
single, larger Boolean algebra of commuting projectors. This is true if and
only if every projector belonging to one of the original algebras commutes with
every projector belonging to the other algebra, which in turn is the same as
requiring that the projectors from the two decompositions of the identity, or
sample spaces, commute with one another.  A larger collection of frameworks is
compatible if all pairs are compatible, and frameworks are said to be mutually
{\it incompatible} if they are not compatible.  Descriptions based upon two or
more compatible frameworks can always be combined by using the single Boolean
algebra which contains all of the different (mutually commuting) algebras, and
thus one is still employing a single framework, corresponding to a single
decomposition of the identity, in accordance with the single-framework rule.

	The single-framework rule is not at all unreasonable from the
perspective of elementary probability theory, where problems are generally set
up using a single sample space.  Thus if a coin is to be tossed ten times in a
row, the statistical properties are worked out not by constructing ten sample
spaces, but by using a single sample space containing $2^{10}$ points.  The
single-framework rule is also perfectly compatible with classical statistical
mechanics, for if one uses two or more coarse grainings of the phase space, the
results can always be combined by means of a single coarse graining which uses
a collection of cells generated by intersections of other cells in an obvious
way.  Thus ordinary probabilistic arguments and classical statistical mechanics
satisfy the single-framework rule, albeit in a somewhat trivial sense.

	As already noted, the single-framework rule as applied to Boolean
algebras of properties refers to a {\it single system} at a {\it single
instant} of time. Given two nominally identical systems, there is no reason why
one cannot use one framework for the first and a different framework for the
second.  For instance, in the case of two spin-half particles, $S_x=+1/2$ could
be a correct description of one of them at the same time that $S_z=+1/2$ is a
correct description of the other.  Similarly, the same particle may have
$S_x=+1/2$ at an earlier and $S_z=+1/2$ at a later time.  Conversely, when
incompatible frameworks turn up in some discussion of a quantum system, it is
best to think of them as referring to different systems, or to a single system
at different times, or perhaps simply as tentative or hypothetical proposals
without any suggestion that they should be taken in a realistic sense.
(Ascribing properties to a single system at more than one time requires the use
of a history, and this requires additional considerations which lie outside the
scope of the present discussion.)

	In any application of probability theory, precisely one of the elements
of the sample space is thought of as existing, or ``true'' in any realization
of an ideal experiment. In this sense the notion of ``truth'' in a
probabilistic theory is necessarily connected with, and thus depends upon the
sample space or framework one is considering.  In classical physics one can
forget about this dependence, because if more than one framework is under
consideration, in the case of a single system at a single time, they can always
be combined into a single framework.  This is reflected in the fact that one
can always define a universal truth functional for a classical phase space, as
noted in Sec.~\ref{s2}.  Because of the possibility that frameworks can be
incompatible, the framework dependence of ``true'' is not at all trivial in
quantum physics; indeed, one might say that this is one of the main ways in
which the mathematical structure of quantum theory forces one to adopt a
different kind of physical interpretation from what one is used to in classical
mechanics.

	In particular, in quantum theory {\it there is no universal truth
functional $\th_q$} which can be used to assign values of 0 and 1 to all
projectors in a way which agrees with the three conditions in (\ref{e.1}). In a
certain sense this is immediately obvious for any Hilbert space of dimension 2
or more, since in such a space there will always be projectors $P$ and $Q$
which do not commute with each other. In such a case $PQ$ is not a projector,
and the third condition in (\ref{e.1}) is not even defined, much less
satisfied.  One might hope to get around this problem by modifying the third
condition and only requiring that it hold in cases in which $P$ and $Q$ commute
with each other.  This, however, gains very little, for the results of Bell and
of Kochen and Specker referred to earlier demonstrate that even such a
``modified'' universal truth functional does not exist for a Hilbert space of
dimension 3 or more.  (A simple example due to Mermin, showing the
impossibility of a universal truth functional in a Hilbert space of dimension
4, is discussed in \cite{gr00a}.)

	The absence of a universal truth functional causes no difficulties for
CH quantum theory because of the single-framework rule, which prevents the
comparison of incompatible frameworks.  The situation is different for the
alternative principle proposed by Bassi and Ghirardi, which they have somehow
managed to confuse with the single-framework rule, and which will be taken up
next.

\section{The Every-Framework Principle of Bassi and Ghirardi}
\label{s4}

	In \cite{bg00} Bassi and Ghirardi introduce, in the discussion leading
up to and including their (6.1), what I shall call the ``every-framework
principle'', which in the notation of the present paper can be stated in the
following way.

	Consider a quantum Hilbert space, and let $\{ \DS_f\}$ be the different
possible decompositions of the identity, where $f$ is a label which takes on
uncountably many values. For example, for a spin-half particle, $f$ will run
over all directions in space $w$, as long as $+w$ is identified with $-w$,
since each decomposition of the identity corresponds to a sample space with
just the two points $S_w=\pm 1/2$.  Corresponding to $\DS_f$ there is a
corresponding Boolean algebra $\BS_f$ of projectors of the form (\ref{e.5}).
Given a projector $P$, we define
\beq
  \FS(P) = \{f: P\in \BS_f\}
\label{e.8}
\eeq
to be the collection of labels such that $P$ is a member of the Boolean algebra
$\BS_f$.

	The {\it every-framework principle} asserts that there is a collection
of truth functionals $\{\th_f\}$, one for each decomposition of the identity,
with the following property: if $P$ is any projector in the Hilbert space, {\it
the value of $\th_f(P)$ is the same for all $f$ in $\FS(P)$}.  That is to say,
$P$ is assigned precisely the same truth value, 0 or 1, by all members of the
collection $\{\th_f\}$ for which $\th_f(P)$ is actually defined.

	The every-framework principle has a certain intuitive appeal when one is
thinking of a single system at a single time.  It is actually correct for
classical statistical mechanics, where  a property $P$ is true as
long as the representative phase point $\gm$ is inside the corresponding set
$\PS$, and false otherwise.  Hence to construct a collection of truth
functionals, associated with a collection of coarse grainings, satisfying the
every-framework principle, one simply chooses some representative point
$\gm_q$ in the phase space, and for a coarse graining $\DS_f$ lets the
indicator for the cell which contains $\gm_q$ play the role of the special
$D_k$ in (\ref{e.6}).  Or, to put the matter in a slightly different way, one
simply lets $\th_f$ be the restriction to $\BS_f$ of the universal truth
functional defined in (\ref{e.7}).

	Given this result, one is not surprised to learn that the
quantum-mechanical version of the every-framework principle implies the
existence of a universal truth functional $\th_u$ of the modified form
discussed towards the end of Sec.~\ref{s3}.  For each $P$, one sets $\th_u(P)$
equal to the common value specified by the every-framework principle, noting
that every $P$ is contained in at least one decomposition of the identity, that
consisting of $P$ and $I-P$.  Since when restricted to the Boolean algebra
$\BS_f$ the functional $\th_u$ is identical to $\th_f$, it is at once evident
that the first two requirements in (\ref{e.1}) will always be satisfied, while
the third will be satisfied in cases in which $P$ and $Q$ commute, since there
is then at least one Boolean algebra $\BS_g$ which contains both of them, and
$\th_u$ when restricted to $\BS_g$ is the same as the corresponding truth
functional $\th_g$.

	But, as we have already noted, the {\it nonexistence} of a universal
truth functional has been proven mathematically for any Hilbert space of
dimension greater than two.  Consequently, {\it the every-framework principle
is in clear contradiction with the principles of quantum mechanics}.  The
proofs of this fact given in \cite{bg99,bg00} are correct but superfluous; they
simply repeat what is already well known to people who work in the foundations
of quantum theory.

	What is the relationship of the every-framework principle and the
single-framework rule?  They are mutually contradictory, for fairly obvious
reasons. The every-framework principle requires us to compare the results of
truth functionals, or equivalently sample spaces, associated with different and
in general incompatible frameworks, in precisely the manner forbidden by the
single-framework rule.  According to the latter, such a comparison makes no
sense in the case of incompatible frameworks.  (Note that it is only with the
help of incompatible frameworks that one can reach a contradiction using the
Bell-Kochen-Specker approach, so incompatible frameworks are essential to the
argument in \cite{bg00}.)  To be sure, two different frameworks might refer to
two different systems (or the same system at two different times), but in that
case there is no reason whatsoever to expect that a particular property has the
same truth value for the two systems, and thus no motivation for invoking the
every-framework principle.

	One must admit that the every-framework principle has a certain
intuitive appeal: how could the truth value of some physical property possible
depend upon the sample space in which it is embedded?  Surely if it is true it
is really true, apart from anything one can say about the rest of the world,
and if it is false it is false!  This appeal is seductive because it focuses
attention on the {\it physical property} rather than on the {\it sample space}.
When, however, one pays attention to the latter, things appear in a quite
different light.  Let us consider as an example two {\it incompatible} quantum
sample spaces
\beq
  \SS_1 = \{A, B, C\}.\quad \SS_2=\{A, D, E\},
\label{e.9}
\eeq
where $A$, $B$, and $C$ are three projectors which add up to $I$, and likewise
$A+D+E=I$.  However, neither $B$ nor $C$ commutes with either $D$ or $E$.

	Let us employ the every-framework principle, and suppose that $A$ is
{\it false} in both $\SS_1$ and $\SS_2$.  Because $\SS_1$ is a sample space,
this means that either $B$ or $C$ is true, and because $\SS_2$ is a sample
space, either $D$ or $E$ must be true.  Suppose for the sake of argument that
it is $B$ which is true in $\SS_1$ and $D$ which is true in $\SS_2$.  Then if
we insist that $\SS_1$ and $\SS_2$ apply to the same system at the same time,
this means that two properties represented by {\it non-commuting projectors}
are simultaneously true.  For example, they could be $S_x=+1/2$ and $S_z=+1/2$
for a spin-half particle.  This is hard to reconcile with the Hilbert space
structure of ordinary quantum mechanics, as pointed out in Sec.~\ref{s3}, and
in CH quantum theory it is forbidden by the single-framework rule.  Thus we see
that when $A$ is false, the every-framework principle has certain implications
which, when brought to light, make it much less appealing.

	The case in which $A$ is {\it true} in both $\SS_1$ and $\SS_2$ also
leads to unsatisfactory results.  Because $\SS_1$ and $\SS_2$ are sample
spaces, the truth of $A$ means that all four properties $B$, $C$, $D$, and $E$
are false.  One might be tempted to suppose that the falsity of two
incompatible properties is unproblematical---after all, who cares about things
which do not occur?  The trouble is that when $B$ is false, its negation
$\tilde B=I-B$ is true. Furthermore, if two projectors $B$ and $E$ do not
commute, the same is true of their negations $\tilde B$ and $\tilde E=I-E$.
Thus using the every-framework principle once again leads to the conclusion
that two properties represented by non-commuting operators are simultaneously
true.  In summary, whatever may be its initial intuitive appeal, much of the
allure of the every-framework principle vanishes when one realizes what it
really means.

	Let us look at this example from a slightly different perspective, by
introducing a third sample space
\beq
  \SS_0 = \{A, \tilde A=I-A\}
\label{e.10}
\eeq
containing only $A$ and its negation.  This is obviously the smallest sample
space in which ``$A$ is true'' and ``$A$ is false'' make sense.  The sample
space $\SS_0$ is compatible with both $\SS_1$ and with $\SS_2$, each of which
represents a {\it refinement} of $\SS_0$.  The rules of quantum reasoning
employed in \cite{gr96} allow one to deduce that if $A$ is true/false in
$\SS_0$, then it is also true/false in $\SS_1$, and the same deduction is
possible going from $\SS_0$ to $\SS_2$.  However, the single-framework rule
prevents combining these results in a single description: one cannot employ
both $\SS_1$ and $\SS_2$ for the same system at the same time, for the reasons
indicated previously.  This means that at least some of the intuitive appeal
which seems to lie behind the every-framework principle, the notion that the
truth of some property should not depend upon what else is going on in the
world, is supported in CH quantum theory.  And this can be done in a consistent
way without leading to any logical contradictions precisely because the CH
approach employs the single-framework rule rather than the every-framework
principle.

	The single-framework rule and the every-framework principle are, thus,
completely incompatible with each other, whether one regards them either from a
purely formal perspective---the former forbids combinations which the latter
allows---or in terms of their intuitive significance. Hence one can only regard
with astonishment the claim of Bassi and Ghirardi, found in the very next
sentence after their (6.1), that the every-framework principle constitutes the
``only reasonable way'' to interpret the single-framework rule!  It is hard to
imagine a more serious misunderstanding of a rule that has been stated over and
over again in the literature on CH quantum theory, and illustrated by means of
numerous examples.  Discussions and criticisms of the single-framework rule can
be a valuable component of the scientific enterprise.  But to introduce a new
principle which is not only different from, but directly contrary to the
single-framework rule, and then claim that the former is the only reasonable
way to interpret the latter does nothing but cause confusion.

	\section{Some Other Issues}
\label{s5}

	Aside from the every-framework principle, there are some other points
in \cite{bg00} (and also \cite{bg00b}) which merit at least a brief response.

	\bt In Secs. 4 and 9 of \cite{bg00}, Bassi and Ghirardi assert that
what I call MQS (macroscopic quantum superposition) states---for example,
Schr\"odinger's infamous cat---are physically unacceptable, and fault the CH
approach for not providing some criterion for excluding them.

	In response, it will help to use an analogy from classical physics,
while remembering that any classical analogy can only go part way in helping us
understand quantum phenomena.  A coin spontaneously rising a centimeter above a
table on which it is sitting at rest is physically unacceptable in the sense
that such a violation of the second law of thermodynamics is never observed to
occur, despite the fact that nothing in the laws of classical mechanics
excludes such a possibility.  We understand why we never observe such things by
using statistical mechanics, which assigns an extremely small probability to
such an event.  That is, we have a scientific understanding of why violations
of the second law are not observed, despite the fact that the laws of classical
(and also quantum) mechanics permit such possibilities.

	The quantum Hilbert space certainly contains MQS states, because it is,
by definition, a linear vector space which includes superpositions of any of
its elements.  However, MQS states are incompatible, in the quantum sense, with
the quasi-classical framework(s) \cite{gmh} needed to describe our
ordinary experience with macroscopic objects.  Thus the single-framework rule
tells us that it makes no sense trying to include MQS states in descriptions of
the everyday world of human experience.  Conversely, if a quantum description
employs the MQS state that is (formally) a linear superposition of a live and
dead cat, it makes no sense, according to the single-framework rule, to think
of the whatever-it-is as somehow involving a cat, for the properties typically
used to identify a cat will, in quantum theory, be represented by projectors
which do not commute with the MQS state, and are therefore of no use for
discussing the meaning of such a state.  In this sense, at least, CH quantum
theory does provide criteria for excluding MQS states from certain types of
quantum descriptions.  Quantum physicists who refuse to employ the
single-framework rule must, of course, find some other means of disposing of,
or perhaps peacefully coexisting with MQS states.  It is also worth noting that
the reason quantum superpositions states of this sort cannot be detected in the
laboratory, even for microscopic objects, as long as they contain a substantial
number of atoms, is by now reasonably well understood in terms of the process
of decoherence, a topic which has been treated from the CH perspective by
Omn\`es \cite{om99}.  (Decoherence is much like classical irreversibility,
making the jumping coin an even better analogy.) To summarize the situation, CH
quantum theory certainly permits descriptions using MQS states, but at the same
time provides an explanation as to why they are neither needed nor particularly
useful for a science of the macroscopic world.

	\bt In a not unrelated point, Bassi and Ghirardi suggest that one may
be able to get around the difficulties they have encountered by employing their
every-framework rule by making a drastic reduction in the set of consistent
families which can be considered to be physically significant. 

	In response, there is nothing wrong with these authors announcing a
direction for their future research, as long as they make it plain that the
motivation for it comes not from any problem involving the CH approach, but
rather from the disagreement between their every-framework principle (itself
completely contrary to the CH single-framework rule) and standard quantum
theory.  No proposal for using a restricted class of families in the manner
they propose has thus far turned out to be very useful for quantum
interpretation, but no doubt those who consider this a worthwhile approach will
continue the search.

	\bt Bassi and Ghirardi take the position, both in Sec.~6 of
\cite{bg00} and in \cite{bg00b}, that the only alternative to their
every-framework principle in which any property $P$ has the precisely the same
truth value in every framework which contains it, is to suppose that in certain
of frameworks it is true and in other frameworks it is false, something they
consider unacceptable.

	The response to this is contained in the material in Sec.~\ref{s4}
above, but it may be worthwhile making it quite explicit.  From the CH
perspective, using two incompatible frameworks to describe the same system at
the same time is not meaningful---this is precisely the point of the
single-family rule.  Since meaningless truth values are meaningless, there is
no reason to be concerned about whether they agree or disagree.  Alternatively,
one can suppose that two incompatible frameworks do {\it not} refer to the same
system at the same time.  In that case, there is no a priori reason to expect
the truth values for a particular property to be the same, and so no reason to
be worried if they are different.

	\bt In \cite{bg00b} Bassi and Ghirardi assert that in the previous
literature on consistent histories the single-framework rule was not explained
well enough or clearly enough so as to obviously exclude their every-framework
principle.

	It is quite true that the language of truth functionals was not
employed by consistent historians (so far as I am aware) prior  to the recent
\cite{gr00a}.  Previous work used the standard language of elementary
probability theory, with its sample spaces and event algebras, and assumed the
usual association between probability theory and reality, as pointed out, for
example, in Sec.~7.2.3 of \cite{gr84}.  A basic understanding of how sample
spaces function in ordinary probability theory is all that one really needs in
order to understand the CH approach and the significance of the
single-framework rule, including the fact that it is quite contrary to the
every-framework principle. The use of truth functionals, while it may be
advantageous for some purposes, is not actually needed.

	\bt At the beginning of Sec.~7.1 in \cite{bg00}, Bassi and Ghirardi, in
a footnote, issue a challenge to me and an anonymous referee to identify which
of their four precisely formulated (in their opinion) assumptions are
inconsistent with the CH single-framework rule, and accept the consequences of
this identification. 

	In fact these four assumptions are not precisely formulated, as was
pointed out in \cite{gr00a}. Writing in response to that, Bassi and Ghirardi
\cite{bg00b} have themselves identified their assumption (c) as the one which
is incompatible with the single-framework rule, and I see no reason to dispute
this.   That rejecting their (c)---the every-framework principle---leads to
dire consequences is not true, as should be clear from the discussion in
Sec.~\ref{s4} above.  Instead, it allows a sensible discussion of quantum
properties using consistent quantum principles.

	\section{Conclusion}
\label{s6}

	In \cite{bg00} Bassi and Ghirardi have, in essence, substituted their
every-framework principle for the single-framework rule of CH quantum theory,
and then concluded, correctly, that the every-framework principle makes no
sense in the quantum world.  Their only mistake is in supposing that the
every-framework principle has something to do with CH quantum theory, whereas
in fact the two are directly contrary to each other.  While this error is
easily spotted by someone who is familiar with CH methods, it is nonetheless
regrettable that others less familiar with them have, once again, been given
the mistaken impression that there is something logically unsound, or at least
suspicious, about CH quantum theory.

	To be sure, Bassi and Ghirardi and other critics of CH perform a
valuable function in looking for flaws in this approach.  Their failure (at
least thus far) to find anything wrong with CH, while at the same time
demonstrating that the various alternatives that they propose posses serious
flaws, adds to one's confidence that the CH approach does, in fact, provide a
satisfactory realistic interpretation of quantum theory.

\section*{Acknowledgments}

	 The author is indebted to T. A. Brun and O. Cohen for reading and
commenting on the manuscript.  The research described here was supported by the
National Science Foundation Grant PHY 99-00755.


\end{document}